\documentclass[%
 reprint,
 amsmath,amssymb,
 aps,
]{revtex4-2}

\usepackage{graphicx}
\usepackage{dcolumn}
\usepackage{bm}
\usepackage{diagbox}
\usepackage[dvipsnames]{xcolor}
\usepackage{changepage}

\begin{document}

\preprint{APS/123-QED}

\title{Estimating angular momenta of fission fragments from isomeric yield ratios using TALYS}

\author{Z. Gao}
\email[]{zhihao.gao@physics.uu.se}
\author{A. Solders}
\email[]{andreas.solders@physics.uu.se}
\author{A. Al-Adili}
\author{S. Cannarozzo}
\author{M. Lantz}
\author{S. Pomp}
\author{H. Sj\"ostrand}
\affiliation{%
 Department of Physics and Astronomy, Uppsala University, 75120 Uppsala, Sweden
}%





\date{\today}

\begin{abstract}
\begin{description}
    \item [Background]
    Angular momenta of fission fragments considerably higher than that of the fissioning nucleus have been observed in many experiments, raising the question of how these high angular momenta are generated. J.~Wilson~\textit{et~al.} have proposed a model for the angular momentum as a function of the mass of fission products based on the assumption that the angular momentum is generated from the collective motion of nucleons in the ruptured neck of the fissioning system. This assumption has caused a lot of debate in the community.
        
    \item[Purpose] To estimate the angular momenta of fission fragments based on the observed isomeric yield ratios in 25-MeV proton-induced fission of $^{238}$U.

    \item[Method] A surrogate model of the fission code GEF has been developed to generate properties of primary fission fragments. Based on the excitation energy and angular momentum of fission fragments from GEF, an energy versus angular momentum matrix is reconstructed using a set of parameters. With such matrices as input, the reaction code TALYS is used to calculate the de-excitation of the fission fragments, including the population of the isomers, from which the isomeric yield ratios are obtained.    
    By varying one of the parameters, the root-mean-square angular momentum ($J_{rms}$), which determines the angular momentum distribution of the matrix, $J_{rms}$-dependent isomeric yield ratios are obtained. 
    Considering all primary fission fragments contributing to the isomeric yield ratio for a given fission product, the average angular momentum of those fragments is estimated. 
    
    \item[Results] Data of 31 isomeric yield ratios in 25-MeV proton-induced fission of $^{238}$U were analysed. From the analysis, the average $J_{rms}$, equivalent to average angular momentum $J^{av}$, with uncertainties are obtained for 24 fission products, while in seven cases no conclusive result for the angular momentum could be obtained.
    Furthermore, considering the neutron emissions of the primary fission fragments, the average angular momentum as a function of the average mass number of the primary fission fragment was estimated. 
    
    \item[Conclusion] A mass dependency of the average angular momentum is observed in the proton-induced fission of $^{238}$U. Moreover, the average angular momenta for mass numbers larger than 131 could be fairly well described by the parameterisation proposed by J.~Wilson~\emph{et al}. However, the average angular momenta of $^{130}$Sn and $^{131}$Te can not be described by Wilson's model, which suggests a different lower limit for the validation of the parameterization in the model. In general, higher average angular momenta for A$\geq$ 132 are observed in the present work compared to those from Wilson~\textit{et~al.} This is likely due to the higher excitation energy of the fissioning nuclei in this work.     
    Furthermore, systematic measurements of the average angular momenta of fission products in the symmetric mass region are presented for the first time. In this region, a decreasing trend with mass number is observed, which can not be explained by the proposal in Wilson's paper. Thus, a different mechanism is needed to explain this observation.  
    
\end{description}

\end{abstract}
\maketitle


\section{\label{Intro}Introduction}
Since the discovery of nuclear fission, considerable effort has been made to understand and describe this phenomenon. The scissioning nucleus, being a quantum many-body system, can not easily be described on the microscopic level, although considerable advancements have been made in the past decades~\cite{Bulgac16, Bulgac22}. As a complement, macroscopic and phenomenological models are often used~\cite{Gef, Stetcu14, Talou14, Litaizea15, Verbeke15}. To optimize and validate these models, accurate experimental data from a broad range of fission reactions and observables are needed.
One of the long-standing questions in fission dynamics is the observation of large angular momenta of the fission fragments. Angular momenta of fission fragments are derived from experimental observables via indirect methods~\cite{Wilhelmy72, Tomar88, Abdelrahman87, Naik95, Naik05, Rakopoulus18, Rakopoulus19, Wilson21}, and the results show that the root-mean-square angular momenta typically are around 6~-~7~$\hbar$. This is, in many cases, considerably higher than those of the fissioning nuclei, which raises the question of how these large angular momenta are generated in the fission process. \par 

In a recent article, J.~Wilson~\textit{et~al.}~\cite{Wilson21} have proposed that the angular momenta are generated after scission, from the collective motion of nucleons in the ruptured neck of the fissioning system. This conclusion is based on fission fragments' angular momenta estimated from $\gamma$-ray spectroscopy, which seems to suggest that the angular momentum of a fission fragment is independent of that of the complementary fragment. Wilson~\textit{et~al.} also propose a parameterization of the angular momentum as a function of mass,d for fission products in the low and high mass peaks.

However, competing explanations for the angular momenta observed by Wilson \emph{et al.} exist. For example, Randrup \emph{et al.}~\cite{Randrup21, Randrup22} claim that the angular momentum is generated before scission. This would mean that the angular momenta of a pair of complementary fission fragments are correlated, but that the correlation may vanish during scission. Another explanation for the lack of correlation is put forward by Stetcu \emph{et al.}. They estimate the angular momentum removal by prompt neutrons and $\gamma$-rays to be in the order of 3.5~-~5~$\hbar$~\cite{Stetcu21}. In their opinion, such a wide distribution of angular momentum removal can hide any underlying correlations of fission fragments' angular momenta. Bulgac \emph{et al.} uses a microscopic model to calculate the angular momenta of fission fragments, and suggest that it is the collective spin modes of the fission fragments that dominate the angular momentum generation~\cite{Bulgac22}.

So far, there is no systematic study, and hence no model, for the angular momenta of fission fragments in the symmetric mass region. Hence, more data on the angular momenta of fission fragments, in particular in the symmetric region, are needed.

In nuclear fission, the prompt energy release, in which the fission fragments (FF) emit prompt neutrons and $\gamma$-rays to form fission products (FP), happens on a timescale of 10$^{-14}$~s~\cite{Wagemans}. This makes direct observations of the properties of the fragments, including excitation energies and angular momenta, impossible with current techniques. Hence, different methods have been developed to deduce the angular momenta from other fission observables. \par 

For example, in the Manchester Spin Method (MSM)~\cite{Abdelrahman87}, the angular momentum is derived from the intensity difference between the ingoing and outgoing $\gamma$-ray transitions of one excited state. As a result, the average angular momenta can be deduced from $\gamma$-ray spectroscopy data~\cite{Wilson21}. \par 

In early attempts, a statistical method based on the Rayleigh distribution~\cite{Huizenga60, Vandenbosch60}, used to describe the angular momentum distribution of the excited nucleus, was used to derive the angular momenta from measured isomeric yield ratios~\cite{Wilhelmy72, Tomar88, Naik05, Naik95}.  
The isomeric yield ratio (IYR), in general terms referring to the relative yield of different long-lived states of a FP, is in this context more specifically defined as the relative yield of the high spin state to the total yield of the FP. \par 

The Madland-England model~\cite{Madland77}, which is commonly used in data evaluations to predict IYR that have not yet been measured, is based on the statistical method. In the Madland-England model, the angular momenta, assumed to be the same for all fragments from a specific fission reaction, is related to the IYR through the spin divider $J_c$, which is determined from the spins of the isomeric states. The model ignores angular momentum removal by neutrons and gammas. Although the model can be used to estimate certain trends in the IYRs, it has been shown to be too crude to give reliable results for angular momentum determinations~\cite{Sears21, Gao23}. \par 

The GEneral description of Fission observables (GEF) model is a publicly available, open-source code which models the whole fission process based on nuclear data, while at the same time conserving the fundamental laws of physics~\cite{Gef}. In GEF, a simplified de-excitation process of the primary FFs determines the correlation of the IYR with the angular momenta, using the same assumption as the Madland-England model. While the internal parameters of GEF are optimized to reproduce fission observables from neutron-induced and spontaneous fission, the predictability of observables from other fission reactions, such as proton-induced fission, is less good~\cite{Gao23}. Furthermore, in cases where GEF fails to reproduce experimental observations, it is not straightforward for the user to tune the relevant parameters, for example, to obtain estimates of the angular momentum of the FFs. \par 

Other fission models, such as MCHF~\cite{Stetcu14}, CGMF~\cite{Talou14}, and FIFRELIN~\cite{Litaizea15}, use the Hauser-Feshbach formalism~\cite{Hauser52} to describe the de-excitation of the FFs, linking the angular momenta to the IYRs. However, these models also use assumptions and model parameters to describe the path from compound nuclei to FFs, which are generally not validated against nuclear data from proton-induced fission.


In an alternative approach to link the angular momenta to the measured IYRs, the reaction code TALYS~\cite{Koning23}, which also uses the Hauser-Feshbach formalism to model the deexcitation, has been used to calculate the yields of the isomers starting from different angular momentum distributions~\cite{Adili19, Rakopoulus18, Rakopoulus19}. In these studies, a clear and strong correlation between the angular momentum and IYR has been observed. Based on that correlation, the average root-mean-square angular momentum $J_{rms}$, determining the angular momentum distribution, has been deduced. However, also in this approach, several assumptions and simplifications are made that might affect the results. For example, the contribution from each FF to the production of the isomer is not calculated, and only three FFs, or even fewer, are considered to contribute to the measured isomeric yield ratio. 
Furthermore, the calculated IYR from each FF is compared with the experimental value to deduce the $J_{rms}$, and the fission yields are not considered in calculating the average $J_{rms}$. Also, the excitation energy of the excited FFs is either modelled using only the mean energy or as a Gaussian distribution. In both cases, the excitation energy is assumed to be independent of the angular momentum distribution. This means that when the value of $J_{rms}$ is increased to an extreme value, the population of the FFs might exceed the yrast line, which is unphysical. \par 

In order to have a more realistic description of the angular momentum and energy distributions of the FFs, TALYS has been coupled to GEF. In this approach, TALYS is used to calculate the yield of the isomers, starting with the properties of the FFs extracted from GEF simulations. The results were evaluated by comparing the derived IYR with measured values, and the conclusion was that the angular momenta of the FFs from GEF was underestimated~\cite{Gao23}.



The present work builds on the GEF + TALYS approach by developing a surrogate model for GEF, in which the $J_{rms}$ of the angular momentum distributions for the FFs can be varied without crossing the yrast line into an unphysical region. The goal is to estimate the most likely angular momenta of the FFs contributing to the yields of the isomers in the 31 measured IYRs from 25~MeV proton-induced fission of $^{238}$U~\cite{Gao23}.\par

\section{Method}

\subsection{Surrogate model}
\label{model}

In this study, GEF is used to provide an initial guess of the states of the FFs. The properties of these states are then tuned using a surrogate model, and used as starting conditions of TALYS calculations.

 
In the development of the model, GEF 2023/1.2 was used to provide the excitation energy (\textit{Ex}) and angular momentum (\textit{J}) distributions of the primary FFs in proton-induced fission of $^{238}$U at 25~MeV. 
Based on this GEF data, the energy vs angular momentum matrix (\textit{Ex-J}) of the excited FFs was parameterised. These parameters could then be used to reconstruct the matrix and use it as input to TALYS, in order to derive the IYR. Thanks to the parameterisation, the parameter determining the angular momentum distribution of the input matrix could be tuned to reproduce the measured IYR.  \par


\subsubsection{GEF: \textit{Ex-J} matrix}
\label{excitation}
Due to the high incident proton energy up to three neutrons can be emitted before scission, resulting in four paths to fission according to GEF. This pre-scission neutron emission changes the scissioning nucleus (SN) as well as its excitation energy. Hence, the properties of the primary fragments produced from different scissioning nuclei will be different.  \par 

\begin{figure}
    \centering
    \includegraphics[width=9cm]{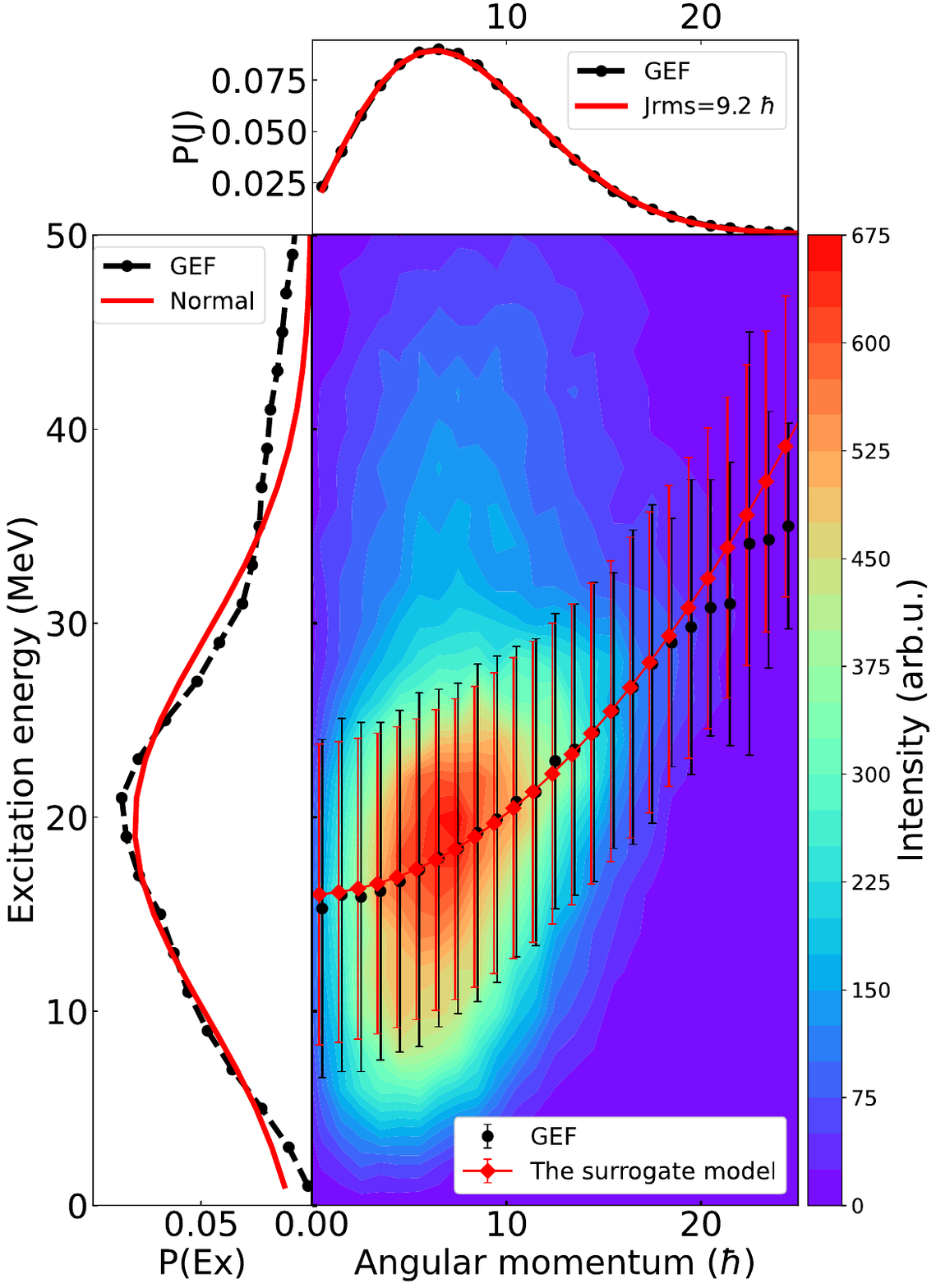}
    \caption{\textit{Ex-J} matrix of the FF $^{133}$Te* produced from the scissioning nucleus $^{237}$Np*, as obtained from GEF, in color-filled contours. In the main panel, black points and bars represent the means and standard deviations of energy distributions for nuclei with a specific angular momentum.
    The red points show a fit to the average energy versus angular momentum. Red bars are the standard deviation of the fitted energies, assumed to be the same for all angular momenta. 
    The left panel shows the excitation energy distribution from GEF in black points and a fitted normal distribution, N(19.6~MeV, 9.6~MeV), in red. The upper panel shows the angular momentum distribution from GEF in black points and a fit of the Rayleigh distribution, giving $J_{rms}=9.2~\hbar$.   }
    \label{133Te_gef}
\end{figure}

As an example, Figure~\ref{133Te_gef} shows the \textit{Ex-J} matrix of the excited FF $^{133}$Te* produced from the scissioning nucleus $^{237}$Np*, as obtained from GEF, in color-filled contours. In the left and upper panels, the matrix is projected to the energy axis and angular momentum axis, respectively. \par 

In GEF, the angular momentum distribution of a particular FF is described by the Rayleigh distribution~\cite{Huizenga60, Vandenbosch60}, 
\begin{equation}\label{rayleigh}
    P(J) = \frac{2\cdot J+1}{2\sigma^{2}} e^{\frac{-J(J+1)}{2\sigma^2}},
\end{equation}
in which the only parameter, the spin cut-off $\sigma$, determines the mean and the width of the distribution. As a result, the root-mean-square angular momentum ($J_{rms}$) from such a distribution is given by, 
\begin{equation}
    J^2_{rms} = 2\sigma^2-\sqrt{0.5\pi}\sigma + 0.25.
\end{equation}
In the literature, this is generally approximated by $J_{rms} \approx \sqrt{2}\sigma$, despite it being off by approximately 0.5~$\hbar$. As seen in the upper panel of Figure~\ref{133Te_gef}, the angular momentum distribution is well described by a Rayleigh distribution with $J_{rms}= 9.2~\hbar$. \par 

The energy distribution, shown in black points in the left panel of Figure~\ref{133Te_gef}, is reasonably well described by a normal distribution N(19.6 MeV, 9.6 MeV), except for energies above 30~MeV. Such a distribution is good enough to serve as a surrogate model. 

If, instead, the data is analysed for one value of the angular momentum at a time, the resulting energy distributions are well described by normal distributions. The means and standard deviations of these distributions are presented as black points with error bars in the main panel of Figure~\ref{133Te_gef}. It can be noticed that the standard deviations of the individual distributions are approximately the same, and are hence replaced by their average value $\bar{\sigma}_{E}$.


From Figure~\ref{133Te_gef}, it is observed that the mean energy for each angular momenta ($\bar{E}_{ex}$) correlates with the angular momentum. Following the parameterisation suggested by Al-Adili~\emph{et~al.}~\cite{Ali21} the relations ship can be described by 
\begin{equation}\label{curvature}
    \bar{E}_{ex}(J) = C \cdot J(J+1) + \bar{E}_{offset} ,
\end{equation} 
in which $C$ and $\bar{E}_{offset}$ are free parameters. In the case of $^{133}$Te, the best fit to the data is $C~=~0.037$~MeV/$\hbar^2$ and $\bar{E}_{offset}~=~16.0$ MeV. 



To summarise, a set of four parameters: $J_{rms}$, $C$, $\bar{E}_{offset}$, and $\bar{\sigma}_{E}$, are needed to describe the \textit{Ex-J} matrix of a single FF from one particular scissioning nucleus.
Considering the four possible scissioning nuclei in the proton induced fission of $^{238}$U at 25~MeV, a total of 20 parameters, adding the four probabilities for each scissioning nuclei, are needed to describe the \textit{Ex-J} matrix. As an example, the values of the 20 parameters describing the four \textit{Ex-J} matrices for $^{133}$Te* are listed in Table~\ref{Parameters}. \par
      
\begin{table}[ht]
    \centering
    \caption{Parameters describing the \textit{Ex-J} matrix of the excited FF $^{133}$Te* obtained from GEF. $P_{SN}$ represents the probability of producing the FF $^{133}$Te* from that particular scissioning nucleus.}
    \begin{tabular}{cccccc}
    \hline
    SN&$P_{SN}$&$J_{rms}$& $C$ &$\bar{E}_{offset}$&$\bar{\sigma_{E}}$ \\
    &&$\hbar$& MeV/$\hbar^2$ &MeV&MeV \\
    \hline
    $^{239}$Np$^*$&0.10&10.8 & 0.025 &35.1& 9.4 \\

    $^{238}$Np$^*$&0.20&10.7 & 0.027 &34.1& 11.2 \\

    $^{237}$Np$^*$&0.37&9.2 & 0.037 &16.0& 7.8 \\

    $^{236}$Np$^*$&0.33&7.6 & 0.036&7.6& 4.9 \\
    \hline
    \end{tabular} 
    \label{Parameters}
\end{table}

\subsubsection{Reconstruction of the \textit{Ex-J} matrix}

The \textit{Ex-J} matrix of any excited FF could be easily reconstructed using the parameters obtained from the GEF data. However, as the aim is to construct a surrogate model in which the angular momentum can be easily varied, having four different values for $J_{rms}$ is cumbersome. 
As a simplification, the four different values of $J_{rms}$ presented in Table~\ref{Parameters} are replaced by a single value. This value is obtained by fitting the total angular momentum distribution, from all scissioning nuclei, with a single Rayleigh distribution.
Together with the values of the other parameters ($C$, $\bar{E}_{offset}$ and $\bar{\sigma_{E}}$ in Table~\ref{Parameters}) this is used to reconstruct the \textit{Ex-J} matrices from the different scissioning nuclei. 
Finally, the four reconstructed matrices are weighted with their respective probabilities ($P_{SN}$ in Table~\ref{Parameters}) and added into one single matrix.

\begin{figure}[ht]
\begin{center}
\includegraphics[width=8.8cm]{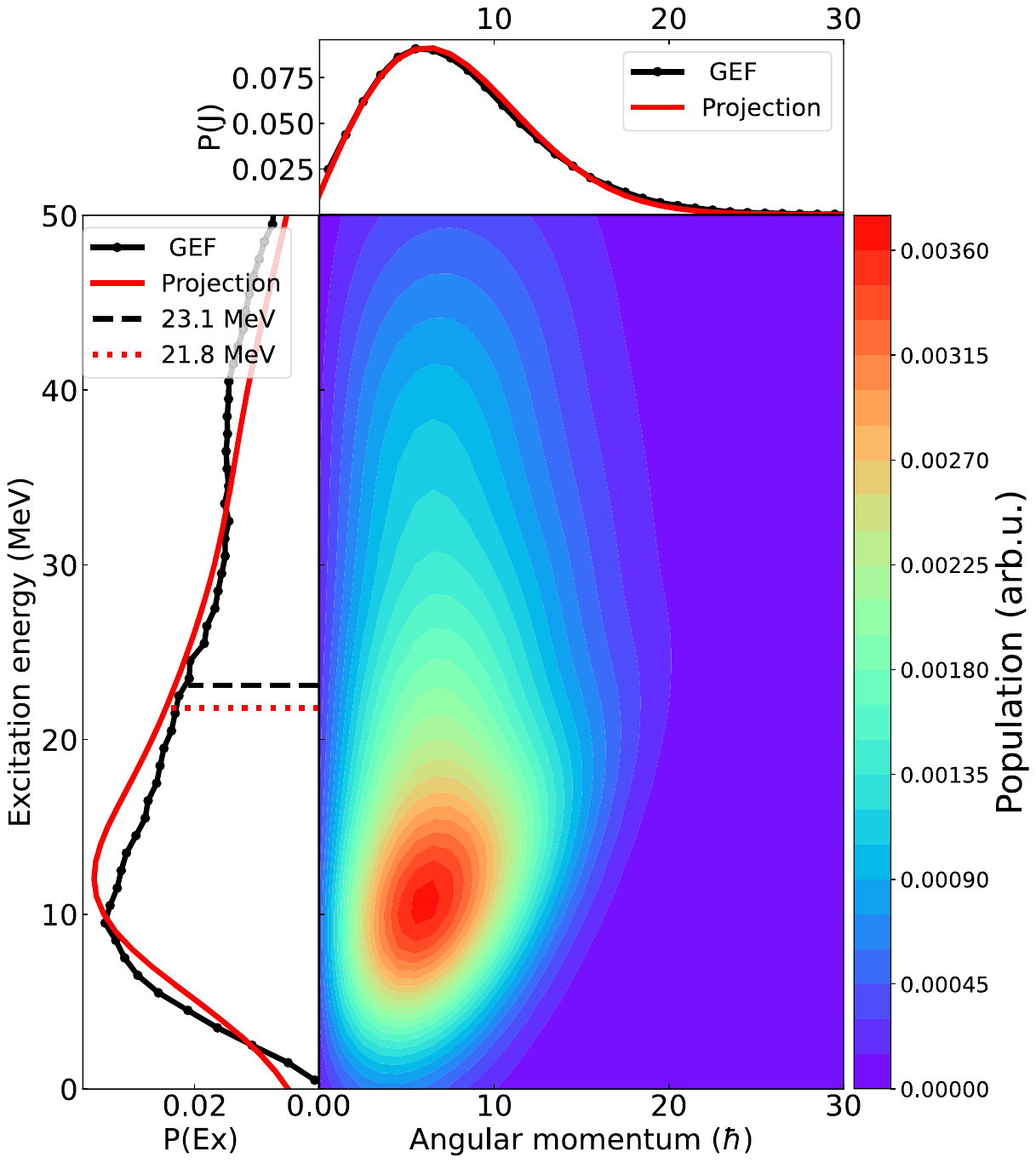}
\caption{Reconstructed \textit{Ex-J} matrix of the primary FF $^{133}$Te, for all four scissioning nuclei. 
The upper and left panels show the angular momentum distribution and the excitation energy distribution from GEF in black dots and corresponding projections of the reconstructed matrix as red curves. The black dashed line, and the red dotted line represent the average excitation energy from GEF and the reconstructed matrix, respectivly.}
\label{recon_m}
\end{center}
\end{figure}

Figure~\ref{recon_m} shows the reconstructed \textit{Ex-J} matrix of the FF $^{133}$Te*, which is a weighted sum of four matrices. Each matrix is reconstructed using a set of three parameters: $C$, $\bar{E}_{offset}$, and $\bar{\sigma}_{E}$ from Table~\ref{Parameters}, and $J_{rms}$~=~9.0~$\hbar$. 
Projections of the matrix to the energy and angular momentum axes are presented in red curves in the left and upper panels, respectively. The corresponding projections of the \textit{Ex-J} matrix from GEF are presented as black lines. The projected angular momentum of the matrix is a good approximation of the GEF data. This means that the single values of $J_{rms}$ could be used to replace the four values in Table~\ref{Parameters}, without serious loss of accuracy. \par 

\begin{figure}[ht]
\begin{center}
\includegraphics[width=8.8cm]{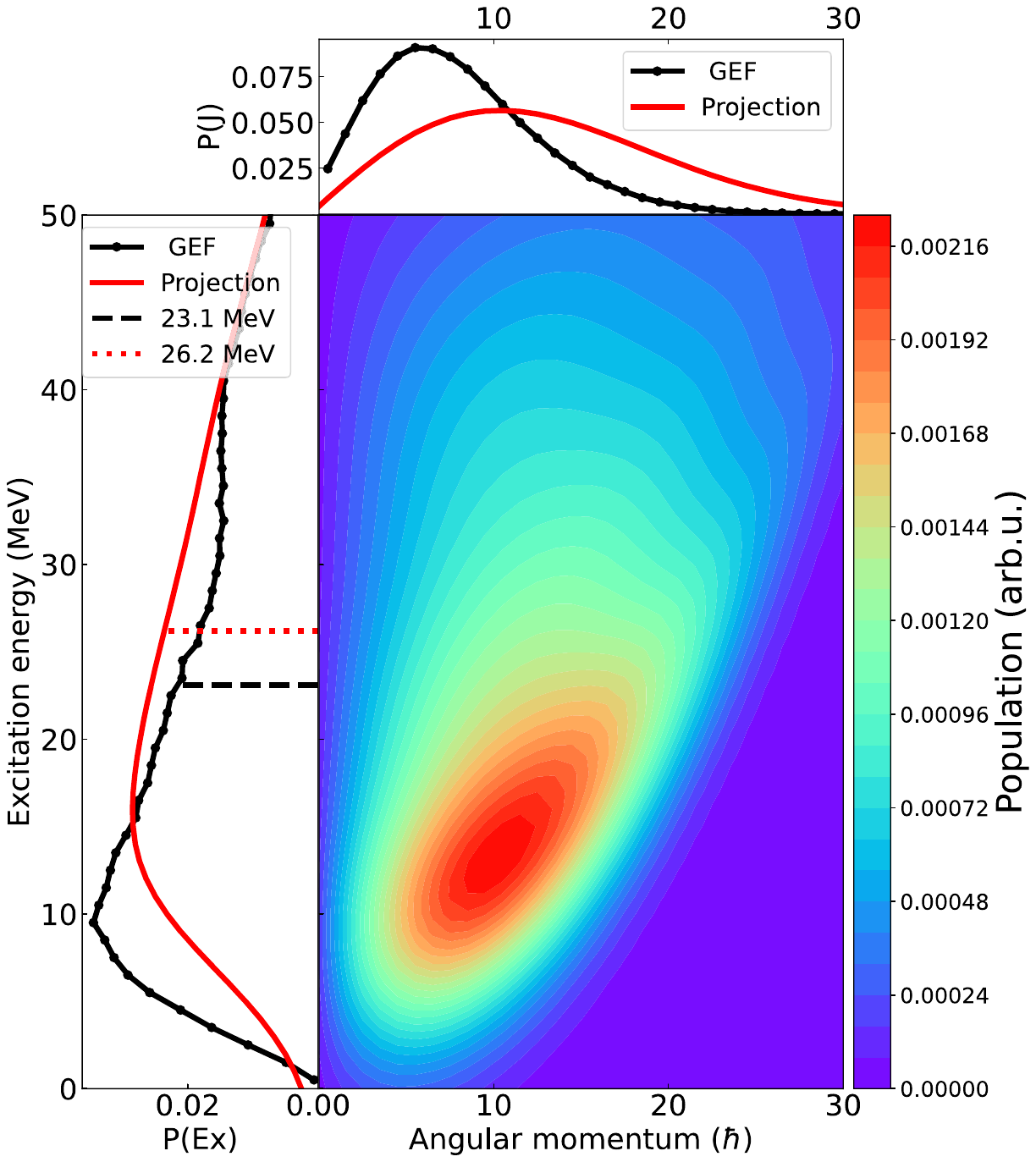}
\caption{Reconstructed \textit{Ex-J} matrix with $J_{rms}= 15.0~\hbar$ for the FF $^{133}$Te* based on the output from GEF. 
The upper and left panels show the angular momentum distribution and the excitation energy distribution from GEF in black dots and corresponding projections of the reconstructed matrix by red curves. The black dashed line, and the red dotted line represent the average excitation energy from GEF and the reconstructed matrix.}
\label{recon_m15}
\end{center}
\end{figure}

To study how the angular momentum affects the de-excitation of the FF, $J_{rms}$ is varied and the resulting matrices are fed to TALYS. As an example, Figure~\ref{recon_m15} shows a reconstructed matrix for the FF $^{133}$Te* with a $J_{rms}$ of $15.0~\hbar$. In the upper panel of Figure~\ref{recon_m15} it can be seen that the reconstructed angular momentum distribution changes as expected. At the same time, the energy distribution is also significantly changed (left panel of Figure~\ref{recon_m15}) due to the dependency on the angular momentum described by Eq~\ref{curvature}.

\subsection{De-excitation with TALYS}
\label{de-ex}

TALYS~\cite{Koning23} is a general nuclear reaction code commonly used in data evaluation. In this study, TALYS 1.96~\cite{talys196}, is used to calculate the de-excitation of excited FFs generated by the surrogate model. \par

The calculations in TALYS were, for each fragment, initialized with a reconstructed \textit{Ex-J} matrix from the surrogate model. However, TALYS is inherently limited to a maximum of 31 discrete values of the angular momenta in the initialisation of the excited nucleus. This means that for large values of $J_{rms}$, the distribution will be cut. 
As an example, in the case of $^{133}$Te, for a $J_{rms}$ of 20~$\hbar$, the integrated probability of spins exceeding the limit ($J~=~30.5~\hbar$) is about 10~$\%$, and as a result, the spin distribution will deviate from the Rayleigh distribution. To avoid large systematic effects of this limitation in TALYS, $J_{rms}$ was limited to a maximum of 25~$\hbar$ in this study. \par

To predict the level densities in the continuum above the largest known discrete energy level, and to supplement the discrete levels whenever the level scheme is incomplete, the Back-shifted Fermi gas Model (BFM)~\cite{Dilg73} was used as the default option.  
In earlier studies, impacts of the choice of level density model on the calculated IYRs have been observed~\cite{Adili19, Rakopoulus18}. Different level density models were therefore tested, and it was concluded that the BFM model led to better agreement with the experimental observations than the other models in most cases. 
However, for cadmium, indium and antimony, the microscopic level densities calculated by S.~Goriely led to better agreement with the experimental ratios and were therefore used instead. \par 

As pointed out by V. Rakopoulos~\emph{et al.}~\cite{Rakopoulus19}, and also observed in the current work, the knowledge of the discrete level scheme can have a significant impact on the calculation of the IYRs in TALYS. Hence, the level schemes used in the calculations were updated with the latest experimental data retrieved from the RIPL-3 database~\cite{Capote09}. In all calculations, TALYS was set up to use all experimentally known levels up to the fortieth level, which is the upper limit in TALYS.





\begin{figure}[ht]
    \centering
    \includegraphics[width=9cm]{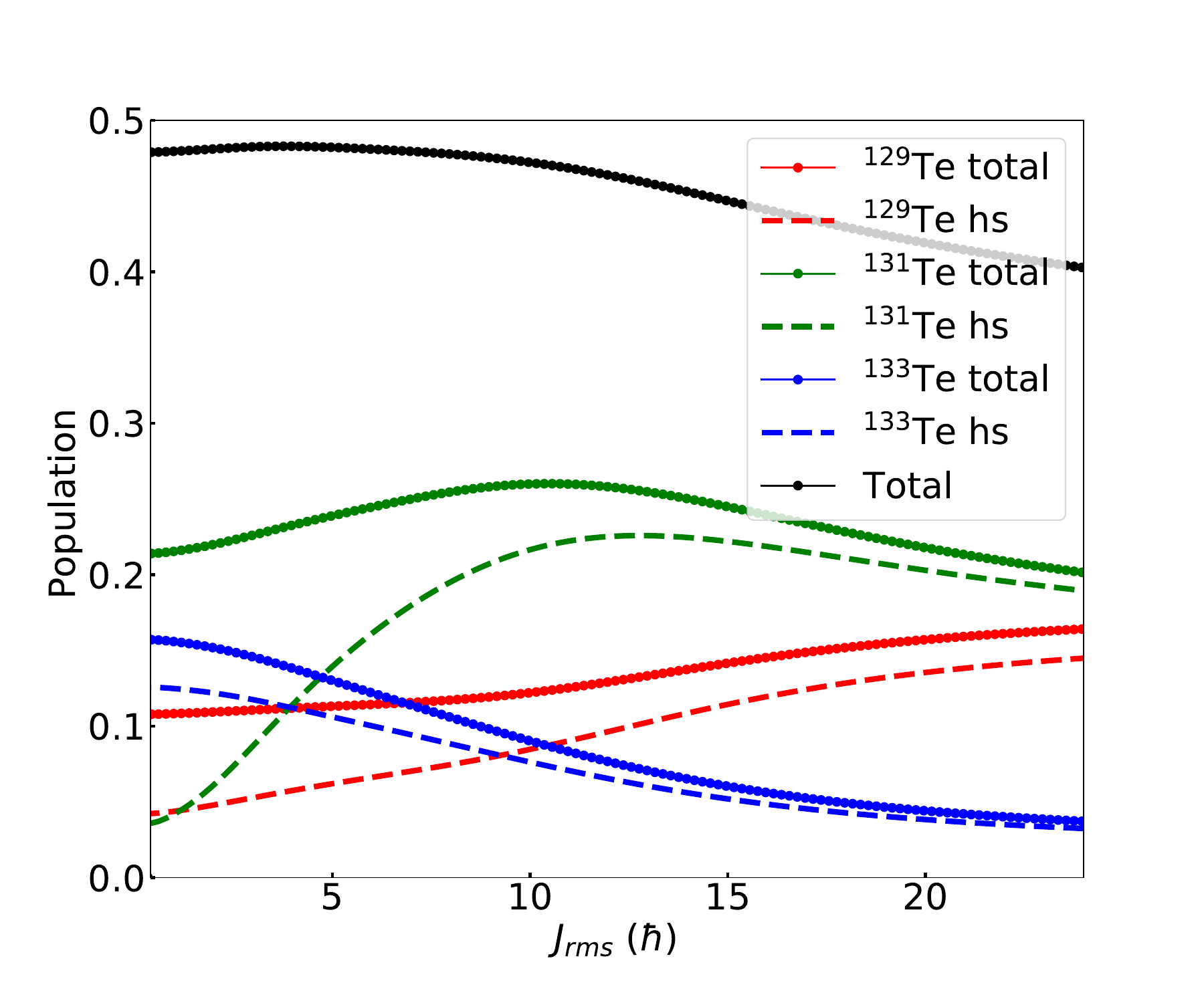}
    \caption{Populations of the FPs from the de-excitation of FF $^{133}$Te* as a function of $J_{rms}$. 
}
    \label{133Te_deex}
\end{figure}

Provided with an input file, TALYS calculates the de-excitation of the excited nucleus and prints the results to an output file. From this output, the population of the long-lived high-spin ($P_{hs}$) and low-spin ($P_{ls}$) states were extracted. For the FF $^{133}$Te*, three of the products after de-excitation, for which experimental data of the IYRs are available for comparison, are $^{129, 131, 133}$Te.  Figure~\ref{133Te_deex} shows the population of these three products as a function of $J_{rms}$ in colored solid curves, while the black curve shows the total population of the three products. 
Considering that the population is normalized to the population of the initial fragment, about half of the FFs $^{133}$Te$^*$ de-excites to either of these three products. 
Hence, observations from these three FPs provide an important part of the information necessary to estimate the $J_{rms}$ of the FF $^{133}$Te*.

Figure~\ref{133Te_deex} also shows that the population of the high spin state has a different dependency on $J_{rms}$ compared to the total population. As a result, the  IYR, representing the relative yield of two states, will depend on the $J_{rms}$. 
This dependency will be discussed further in the next section. \par 

Furthermore, it can be observed in Figure~\ref{133Te_deex} how the summed population of two states of $^{129}$Te increases with $J_{rms}$, while the population of $^{133}$Te decreases. The population of $^{131}$Te increases up to about $10.6~\hbar$, and then starts to decrease. 
Such dependencies are not surprising, considering that the excitation energy of the fragment increases with angular momentum (see Eq.~(\ref{curvature})), leading to larger neutron emission. 


\subsection{Angular momentum}
\label{Derivation}

\begin{figure}[ht]
    \centering
    \includegraphics[width=9.4cm]{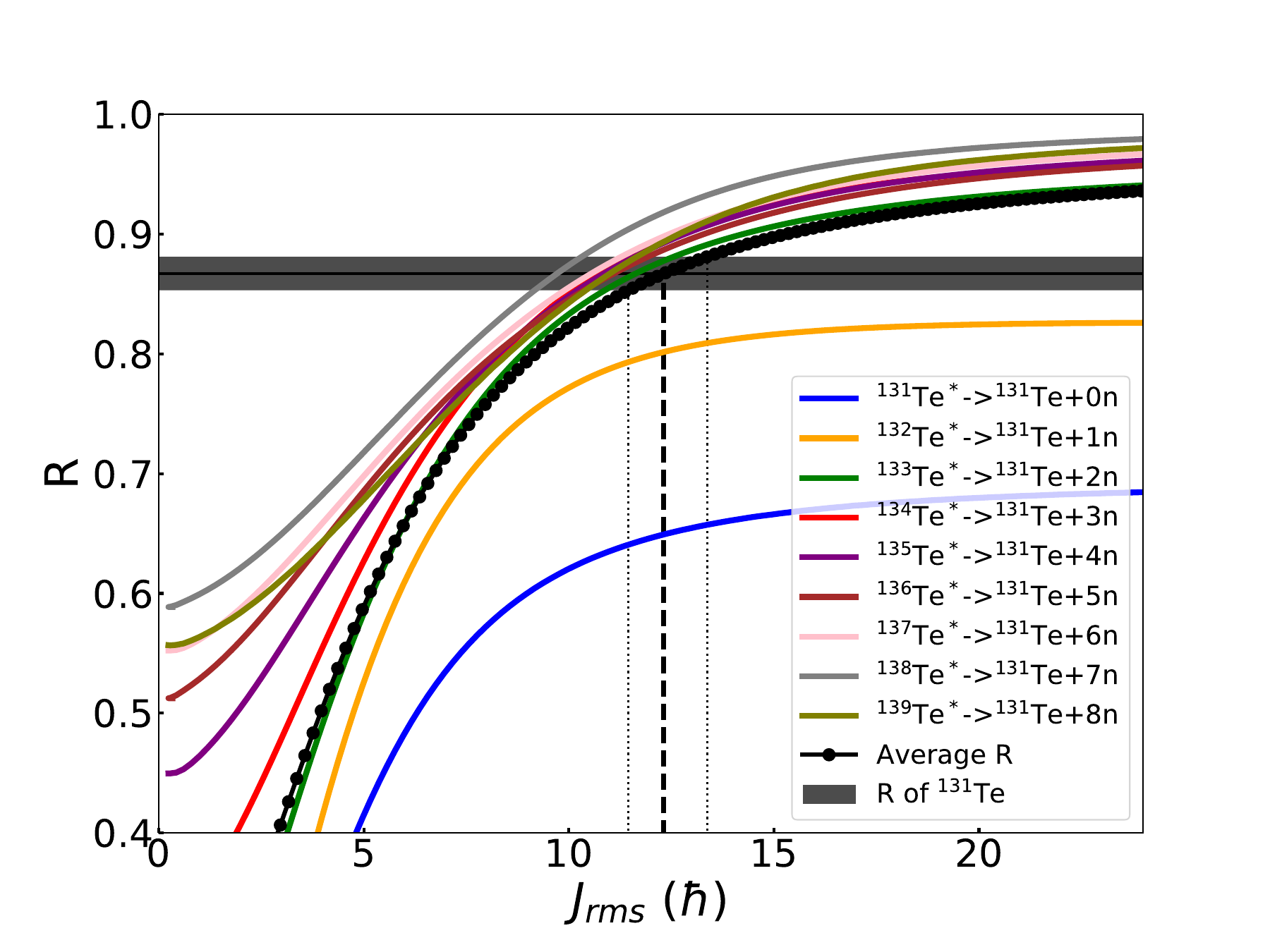}
    \caption{Calculated IYRs of the FP $^{131}$Te from all excited FFs, $^{131-139}$Te*, contributing to the production of $^{131}$Te. The horizontal line and grey region represent the measured IYR of the FP $^{131}$Te with uncertainties~\cite{Gao23}. The vertical dashed line marks the value of J$_{rms}$ where the average curve match the measured ratio.}
    \label{131Te_jrms}
\end{figure}

From the results of the TALYS simulations, the populations of the high spin and low spin states are used to calculate the IYRs, 
\begin{equation}\label{ratio}
    R = \frac{P_{hs}}{P_{hs}+ P_{ls}}.
\end{equation} 
As described above, multiple neutron emission is considered in the de-excitation. Hence, each FP can be produced from multiple excited FFs.
As an example, the derived IYRs of $^{131}$Te, as a function of $J_{rms}$ of the contributing FFs, are presented in Figure~\ref{131Te_jrms}.\par


As seen in Figure~\ref{131Te_jrms}, the calculated IYRs from different initial FFs show a similar dependency on the $J_{rms}$, indicating that common factors affect the calculated IYR. One such factor is the level scheme, including the two long-lived states, of $^{131}$Te. On the other hand, the observed variations between the curves could indicate that the FFs have different impacts on the IYR. \par 

Considering all contributing FFs, the weighted average IYR as a function of $J_{rms}$ was calculated,
\begin{equation}
    \label{w_av_R}
    \bar{R} (J_{rms}) = \frac{\sum^N_{i=0} P_{hs} \cdot Y_i}{\sum^N_{i=0} (P_{hs} + P_{ls})\cdot Y_i },
\end{equation}
where $N$ is the number of possible contributors, and $Y_i$ is the independent fission yield of the contributor, obtained from the GEF simulation. \par 

As an example, the calculated average IYRs of the FP $^{131}$Te are presented in black points in Figure~\ref{131Te_jrms}. As seen, the average IYR matches the experimental ratio at a $J_{rms}$ of $12.3~\hbar$. Furthermore, the calculated IYRs cross the experimental ratio minus (plus) one sigma at a $J_{rms}$ of $11.5~\hbar$ ($13.4~\hbar$). Hence, an average value of $12.3^{1.1}_{0.8} ~\hbar$ is assigned as the most likely $J_{rms}$ of the FFs de-exciting to the FP $^{131}$Te.

\begin{figure}[ht]
    \centering
    \includegraphics[width=9.4cm]{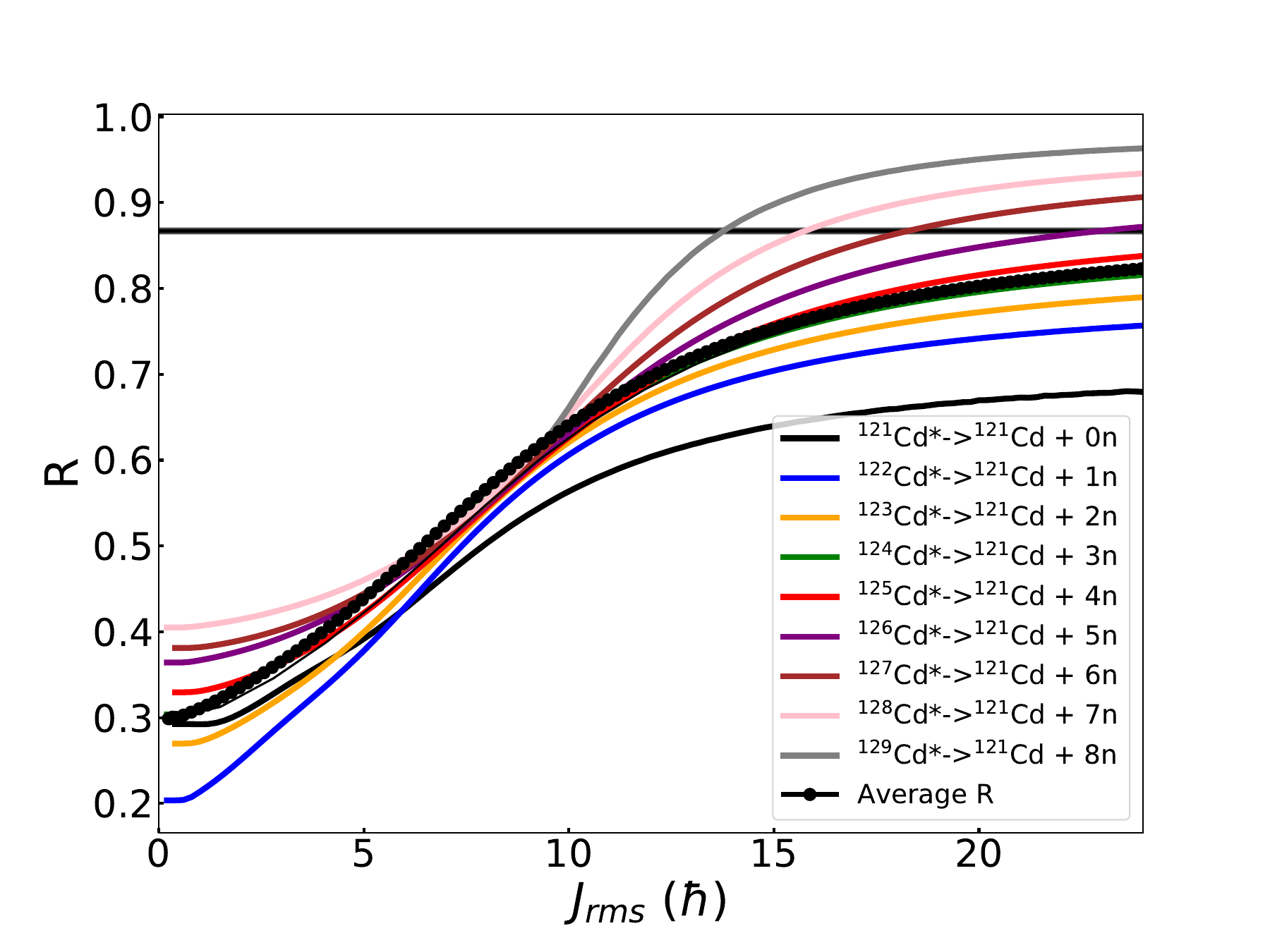}
    \caption{Calculated IYRs of the FP $^{121}$Cd from all excited FFs, $^{121-129}$Cd*, contributing to the production of $^{121}$Cd. The horizontal line and grey region represent the measured IYR of the FP $^{121}$Cd with uncertainties~\cite{Gao23}.}
    \label{121Cd_jrms}
\end{figure}

The procedure described above is repeated for all 31 IYRs. However, in some cases the calculations do not reproduce the experimental value for any $J_{rms}$ in the range 0.2 to 25~$\hbar$. Figure~\ref{121Cd_jrms} shows the example of the IYR of the FP $^{121}$Cd, obtained from the de-excitations of the FFs $^{121-129}$Cd*. As seen, the average IYRs do not reach the experimental ratio for any $J_{rms}$, although some of the ratios for single FFs ($A={126-129}$) can reproduce the experimental ratio. \par  

If we assume that the experimentally determined IYR is correct, there are at least three other possible reasons for the failure of the model to reproduce the experimental data. First of all, besides $J_{rms}$, the values of the other parameters ($C$, $\bar{E}_{offset}$, $\bar{\sigma_{E}}$) also affect the energy and angular momentum population of the FFs. 
In principle, also these parameters could be optimized against experimental data. However, this would require a much larger data set, probably including neutron and gamma multiplicities, in order not to have an underdetermined problem.
Secondly, the de-excitation process, as described by the Hauser-Feshbach approach in TALYS, might need adjustment. Thirdly, and probably the most likely reason, is the lack of experimental excitation levels, and poor knowledge of de-excitation paths toward the long-lived states of the FP. In the case of $^{121}$Cd, the number of known levels in the complete scheme is only 9, indicating a poor knowledge of the nuclear structure. \par 

Furthermore, several studies~\cite{Cannarozzo23, Rodrigo23, Sudar18, Khandaker23} have shown that the width of the angular momentum distribution can have a substantial impact on the calculated isomeric yield ratios for light particle-induced reactions. In TALYS, the width of the predicted spin distribution from the level density models can be controlled by scaling the spin cut-off parameter using a parameter called \textit{Rspincut}.
The studies indicate that the optimal value of the \textit{Rspincut} parameter in light particle-induced reactions is about 0.5, which reduces the width of angular momentum distribution compared to the default value of 1. In this study, this value was tested with the surrogate model, leading to, in most cases, a significantly larger angular momentum than that obtained with an \textit{Rspincut} of 1. However, changing \textit{Rspincut} to 0.5 does not improve the surrogate model's ability to reproduce the experimentally determined IYRs. Since many parameters in the model affect the calculated IYR, and most of them are not yet optimized due to lack of data, the impact of \textit{Rspincut} for this data set could not be determined. Hence, only the results with the default value of \textit{Rspincut} of 1 are presented in this paper.


\subsection{Primary contributors}

As seen in Figure~\ref{131Te_jrms} and~\ref{121Cd_jrms}, the IYR as a function of $J_{rms}$ behaves differently for different FFs. In forming the weighted average ratios in Eq.~(\ref{w_av_R}) the relative contribution from each fragment is taken into account. Although all possible contributors are included, the ratio is largely determined by two to four of the contributors. This can be shown by studying the relative contributions, which can be calculated by multiplying the independent yield of each fragment with the fraction of that fragment that decays to the studied FP. The result  for $^{131}$Te and $^{121}$Cd is presented in Table~\ref{131Te_D}.
At the same time, the average number of neutrons emitted from these fragments, when de-exciting to a specific FP, can be calculated.



Four FFs, $^{132-135}$Te*, together contribute to more than 80\% of the production of the FP $^{131}$Te, which is why the curve for the average IYR is close to the corresponding curves in Figure~\ref{131Te_jrms}. For the production of $^{121}$Cd, the primary contributors are $^{123-126}$Cd*.

\begin{table}[ht]
    \centering
    \caption{Contributions to the production of $^{131}$Te and $^{121}$Cd in the TALYS calculations. Contributions larger then 10~\% are marked in bold font.}
    \renewcommand{\arraystretch}{1.5}
    \begin{tabular}{llccccccccc}
    \hline
     
     Te & A&131 &132&133&134&135&136&137&138&139\\
     Contr. & (\%)&1.7 &\textbf{14.6}&\textbf{31.7}&\textbf{25.3}& \textbf{17.1}&7.7&1.9&0.2&0.01 \\
     \hline
     Cd &A &121 &122&123&124&125&126&127&128&129\\
    Contr. &(\%) &0.39 &4.2&\textbf{18.8}&\textbf{37.8}& \textbf{26.5}&\textbf{10.4}&1.7&0.1&0.003 \\
    \hline
    \end{tabular}
    
    \label{131Te_D}
\end{table}

\section{Results and discussion}
When experimental determinations of angular momentum in fission are discussed in the literature, the deduced values are usually associated with the FPs in question, even though it is the angular momentum of the initial fragment that is sought. In this study, we are also unable to associate an estimated $J_{rms}$ to a specific FF, but rather to a range of FF isotopes. With Table~\ref{131Te_D} in mind, the deduced value of $12.3^{1.1}_{0.8} ~\hbar$ can for example be considered a reasonable estimate of the $J_{rms}$ for FFs of Tellurium isotopes in the mass range $132-135$. 

To simplify the discussion and comparison with earlier studies, the results will, in the succeeding, be presented as a function of the FP for which the IYR is measured. Using the surrogate model and experimental values on the IYR~\cite{Rakopoulus18, Rakopoulus19, Gao23}, the average $J_{rms}$ have been estimated for 31 FPs in proton-induced fission of $^{238}$U, and the results are presented for each FP in Table~\ref{J_FP}.

As mentioned in the introduction, the approach to derive $J_{rms}$ described in this study builds on earlier work. To compare the methods, the 10 values of $J_{rms}$ presented by V.~Rakopoulos \emph{et al.}~\cite{Rakopoulus18, Rakopoulus19} are also presented in Table~\ref{J_FP}.


\begin{table}[!ht]
    \centering
    \caption{Estimateed values of $J_{rms}$ with lower and upper limits from the experimental IYRs in the proton-induced fission of $^{238}$U at 25 MeV~\cite{Rakopoulus18, Rakopoulus19, Gao23}. Blank positions mean that the calculations could not reproduce the experimental values for any $J_{rms}$ below 25~$\hbar$. The last column is the values reported by V.~Rakopoulos \emph{et al.}}
    \renewcommand{\arraystretch}{1.2}
    \begin{tabular}{cccccl}
    \hline
        Nuclide~~ &IYR &$J^{av}_{rms}$ & Lower  & Upper  & $J^{av}_{rms}$~($\hbar$)  \\
        
        &~\cite{Rakopoulus18, Rakopoulus19, Gao23}&($\hbar$)&($\hbar$)&($\hbar$)&~\cite{Rakopoulus18,Rakopoulus19} \\ \hline
        $^{102}$Nb &0.735(21)& &  &  &   \\ 
        $^{119}$Cd &0.871(15)& 15.7 & 1.4 & 1.9 & 12.3(5)  \\ 
        $^{121}$Cd &0.867(4)& ~ & ~ & ~ & 14.7(1)  \\ 
        $^{123}$Cd &0.876(7)& 12.3 & 0.6 & 0.7 & 15.7(2)  \\ 
        $^{125}$Cd &0.902(8)& 10.4 & 0.6 & 0.7 &   \\ 
        $^{127}$Cd &0.87(4)& 12 & 3 & 12&   \\ 
        $^{119}$In &0.978(15)& 22.3 & 6.9 & 2.1 & 26.4(4)  \\ 
        $^{121}$In &0.971(11)& ~ & ~ & ~ & 25.1(5)  \\ 
        $^{123}$In &0.958(2)& ~ & ~ & ~ & 21.2(2)  \\ 
        $^{125}$In &0.950(3)& 12.7 & 0.6 & 0.7 & 15.9(9)  \\ 
        $^{126}$In &0.574(16)& 8.6 & 0.3 &0.3  &   \\ 
        $^{127}$In &0.921(2)& 12.9 & 0.3 & 0.3 & 9.5(2)  \\ 
        $^{128}$In &0.58(4)& ~ & ~ & ~ &   \\ 
        $^{129}$In &0.884(8)& ~ & ~ & ~ &   \\ 
        $^{128}$Sn &0.580(20)& 11.1 & 0.5 & 0.5 & 7.9(4)  \\ 
        $^{129}$Sn &0.777(20)& 8.0 & 0.5 & 0.7 &   \\ 
        $^{130}$Sn &0.540(20)& 10.8 & 0.9 & 1.2 & 6.4(2)  \\ 
        $^{131}$Sn &0.681(12)&   &   &   &   \\ 
        $^{129}$Sb &0.539(21)&10.4 & 0.4 & 0.4 &  \\ 
        $^{132}$Sb &0.433(29)& 8.5 & 0.5 & 0.6 &   \\ 
        $^{134}$Sb &0.625(10)& 7.6 & 0.4 & 0.40 &   \\ 
        $^{129}$Te &0.832(4)& 23.3 & 0.9 & 1.0 &   \\ 
        $^{131}$Te &0.867(14)& 12.3 & 0.8 & 1.1 &   \\ 
        $^{133}$Te &0.794(2)& 8.39 & 0.06 & 0.06 &   \\ 
        $^{132}$I &0.542(20)& 8.2 & 0.3 & 0.3  & ~ \\ 
        $^{133}$I &0.27(3)& 7.3 & 0.4 & 0.4  & ~ \\ 
        $^{134}$I &0.639(5)& 9.4 & 0.1 & 0.1  & ~ \\ 
        $^{136}$I &0.730(20)& 10.3 & 1.0 & 2.2  & ~ \\ 
        $^{133}$Xe &0.618(13)& 5.8 & 0.2 & 0.2  & ~ \\ 
        $^{135}$Xe &0.729(7)& 8.2 & 0.2 & 0.2  & ~ \\ 
        $^{138}$Cs &0.799(17)& 8.7 & 0.5 & 0.6  & ~ \\  \hline
    \end{tabular}
    \label{J_FP}
\end{table}

As mentioned in Section~\ref{de-ex}, there is a limit in TALYS of 30 discrete spin values. Hence,  if a FF is initialized with a $J_{rms}$ larger than about $20~\hbar$ the angular momentum distribution will be significantly cut. As a result, values of $J_{rms}$ larger than $20~\hbar$ should be considered less trustworthy.

\subsection{Comparison of methods}


Although there are many similarities between the method described by V. Rakopoulos \emph{et al.}~\cite{Rakopoulus18, Rakopoulus19} and the one presented in this paper, there are also important differences.
In the method of Rakopoulos, the excitation energy of the FFs is described by a single mean energy with a width of 0.5~MeV, instead of the more realistic energy distribution used in this study. This simplification also ignores the correlation between the excitation energy and angular momentum, here described by Eq~(\ref{curvature}), and as a result, the FF population might extend past the yrast line.
Furthermore, multi-fission channels were ignored and
the yields of the FFs were not considered in the calculation of the weighted $J_{rms}$ of the FFs.\par

Another difference between the two models is that the previous method uses the Skyrme-Hartree-Fock-Bogoluybov table, containing microscopic level densities, to calculate the IYRs of the cadmium, indium and tin isotopes. In this study, the back-shifted Fermi gas model (BFM) is instead adopted, as it was shown to have the best overall performance. 


Due to the aforementioned differences between the two methods, it is not surprising that discrepancies are observed in Table~\ref{J_FP}. 
In a few cases ($^{121}$Cd and $^{121, 123}$In), the model by Rakopoulos could reproduce the measured IYR, while the present method fails. The reasons for this have been discussed in section~\ref{de-ex}. For $^{121}$Cd, the IYR from the surrogate model could be increased by reducing the value of parameters $C$ and $\bar{\sigma_{E}}$ in the reconstruction of the excitation matrices, leading to a match between the calculated and measured IYR.
On the other hand, in the case of $^{121, 123}$In it does not matter how much the other parameters of the surrogate model are varied, the resulting IYRs will not match the experimental IYR for any $J_{rms}$. The reason could be in the assumptions of the model, for example, in how the matrix is reconstructed. However, before drawing any conclusions, the level schemes of $^{121, 123}$In should be carefully examined.
In the cases of $^{125, 127}$Cd the situation is reversed. Here, the IYRs could not be reproduced by the previous method, while the surrogate model produces $J_{rms}$-values, as seen in Table~\ref{J_FP}.

\begin{figure}[ht]
    \centering
    \includegraphics[width=9.4cm]{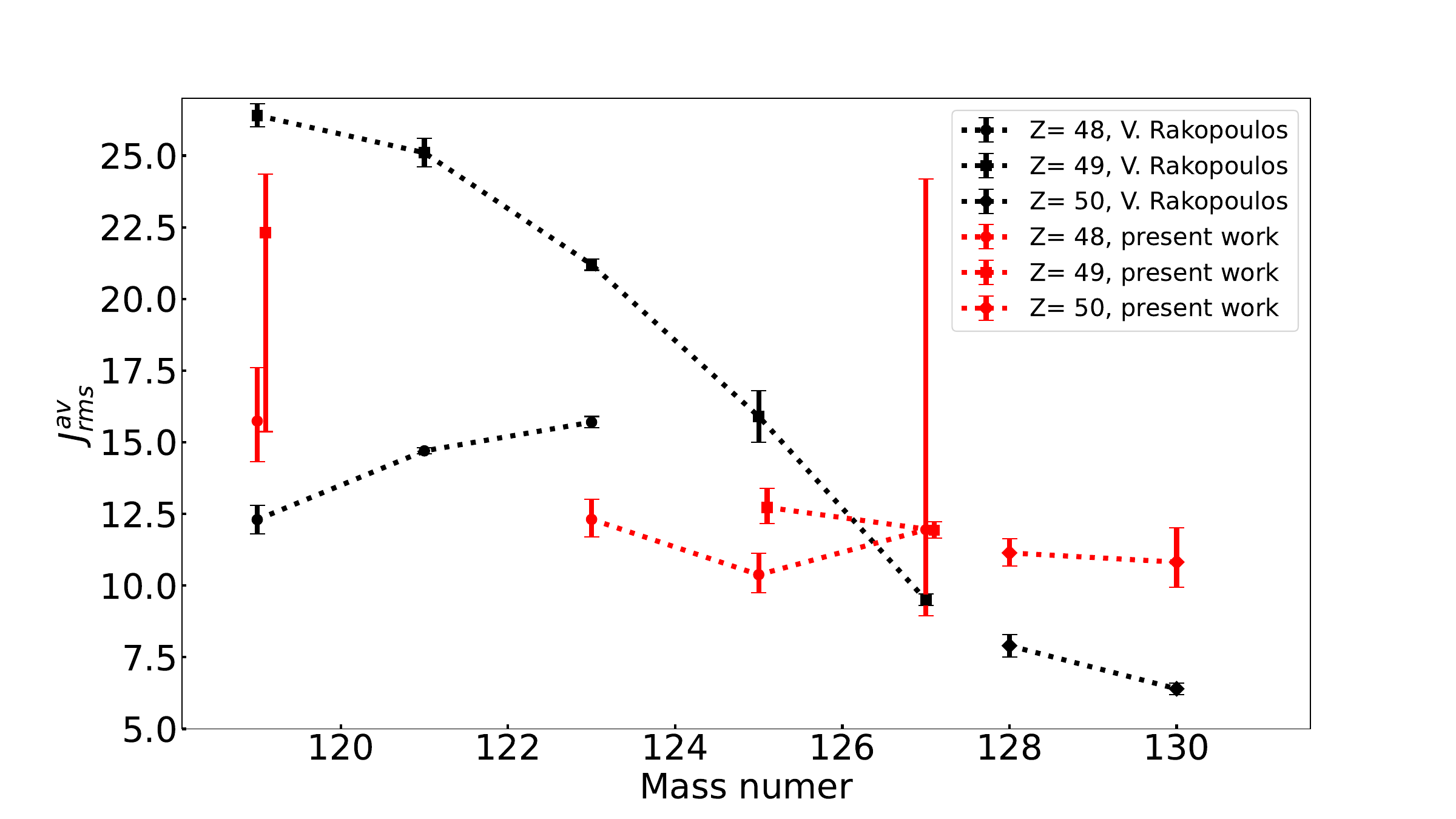}
    \caption{A comparison of the deduced average $J_{rms}$ as a function of mass number for the cadmium, indium and tin isotopes with the two methods. }
    \label{J_bill}
\end{figure}

Figure~\ref{J_bill} shows the derived average $J_{rms}$ as a function of mass number for the measured cadmium, indium and tin isotopes with the two methods. With the surrogate model, the derived average $J_{rms}$ values range from about 10~$\hbar$ to 16~$\hbar$, except for $^{119}$In where the value is larger. With Rakopoulos' method, a larger variation in the average $J_{rms}$ is observed, ranging from 6~$\hbar$ to 26~$\hbar$. \par 

\subsection{Average angular momenta}
\label{Compare_J}
J. Wilson \textit{et~al.} have suggested that the angular momenta of FFs are generated from the collective motion of nucleons in the ruptured neck of the fissioning system~\cite{Wilson21}. Based on the Rayleigh distribution and the Fermi gas model, an equation to describe the observed mass dependency of the average angular momentum is proposed: 
\begin{equation}\label{J_mass}
    J^{av} = cA^{1/4}_{N}A^{7/12}_F,
\end{equation}
in which $c$ is a free parameter and $A_F$ is the fragment mass. $A_{N}$ is a parameter that for FFs near the doubly magic $^{132}$Sn is $A_F - 130$, while for light fragments near the doubly magic $^{78}$Ni it is $A_{F} - 78$. 
In their paper, J. Wilson \textit{et~al.} fit this equation to their data for fast (average energy 1.9~MeV) neutron-induced fission of $^{238}$U. However, they do not present any data, and there is also no description or prediction of the $J^{av}(A)$, in the symmetric mass region.

\begin{figure}
\begin{center}
    \includegraphics[width=9.4cm]{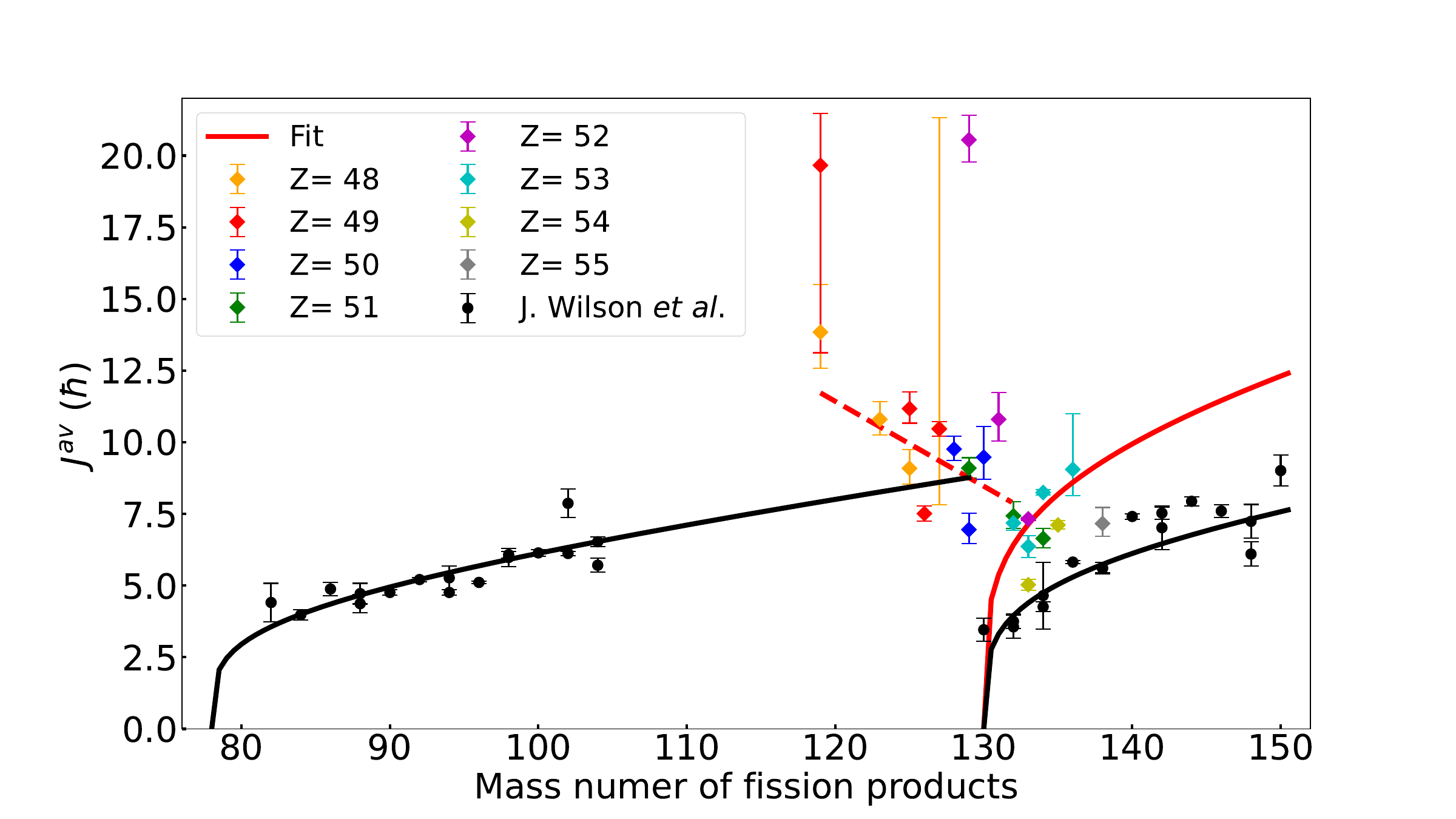}
    \caption{Average angular momentum as a function of mass number of the resulting FP. The colored points present data from proton-induced fission of $^{238}$U, and the red curve is fit to these data for A greater than 131. The red dashed line is obtained from a linear fit to the data for A less than 133. \\
    The black points and curves present the average angular momenta for the FFs in the fast neutron-induced fission of $^{238}$U, and a fit to the data from J. Wilson~\textit{et~al.}~\cite{Wilson21}.}
    \label{J_av}
\end{center}    
\end{figure}

To compare the results of this study with the results by Wilson \textit{et~al.}, the root-mean-square angular momenta ($J_{rms}$) in Table~\ref{J_FP} have been converted to average angular momenta ($J^{av}$). The result is presented in Figure~\ref{J_av} as a function of the mass number of the FP. For comparison, the experimental values and fit from the paper by Wilson \textit{et~al.} are included in the same figure.

Eq.~(\ref{J_mass}) was also fitted to the proton-induced data for A larger than 131, and is shown as a red curve in Figure~\ref{J_av}. The model describes the data rather well. However, the average angular momenta associated to $^{130}$Sn and $^{131}$Te are much higher than the values predicted by Eq.~(\ref{J_mass}). 
 The failure to describe these nuclides might indicate that Eq.~(\ref{J_mass}) is not valid for mass numbers close to 130. This observation is also supported by the fact that the model predicts $J^{av}$ to be zero for A~=~130. On the other hand, it could be argued that the minimum angular momentum should be observed for the most spherical FPs around the doubly magic $^{132}$Sn. \par 

J.~Wilson~\textit{et~al.} do not present any data for the angular momentum of FPs in the symmetric mass region in which mass numbers are close to half of the fissioning nucleus. Our data, however, suggest that the average angular momenta in this region decrease with increasing mass number. A linear fit to the data gives an average decrease by $0.3\pm0.2~\hbar$ per atomic mass unit. This behaviour can not be explained with the model presented by  J. Wilson~\emph{et al.} Since the average angular momentum is assumed to be generated from the collective motion of nucleons in the ruptured neck, the angular momentum must increase with the number of nucleons. Even if one changes the parameters in Eq.~(\ref{J_mass}), it is impossible to predict a decreasing trend of the average angular momentum. Hence,  a different mechanism is needed to explain the observed decreasing trend of the average angular momentum in the symmetry region.



\subsection{Neutron emission}
\begin{figure}[ht]
    \centering
    \includegraphics[width=9.4cm]{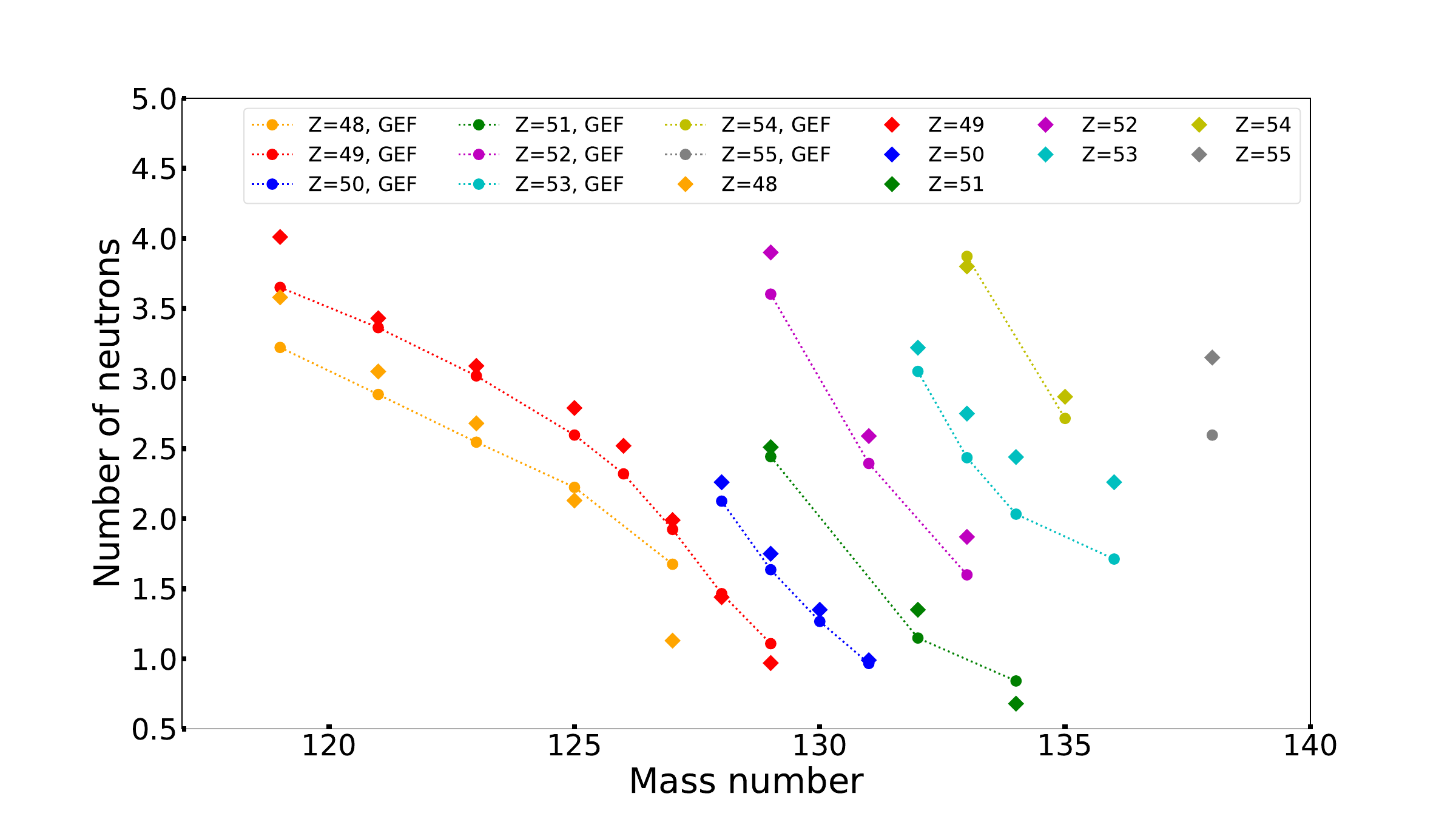}
    \caption{The average number of neutrons emitted from the primary FFs de-exciting to a specific FP as derived from TALYS, compared with the same information from GEF.}
    \label{n_mass}
\end{figure}

From the TALYS output of the surrogate model, information on the neutron multiplicity for each FF can be extracted. From this information, the average neutron multiplicity for each fragment decaying to a certain FP can be derived. In Figure~\ref{n_mass}, this is plotted together with the same information obtained directly from the GEF model, and a similar behaviour of the two models is observed. Although the neutron multiplicity in proton-induced fission in GEF is not validated against experimental data, and should hence not be taken as a benchmark, the agreement builds some confidence in the surrogate model. The discrepancies that can be observed are likely due to the different ways of dealing with the de-excitation process, and different assumptions of angular momenta of the FFs.

 
\subsection{Angular momentum of the primary fission fragment}
\begin{figure}[ht]
    \centering
    \includegraphics[width=9.4cm]{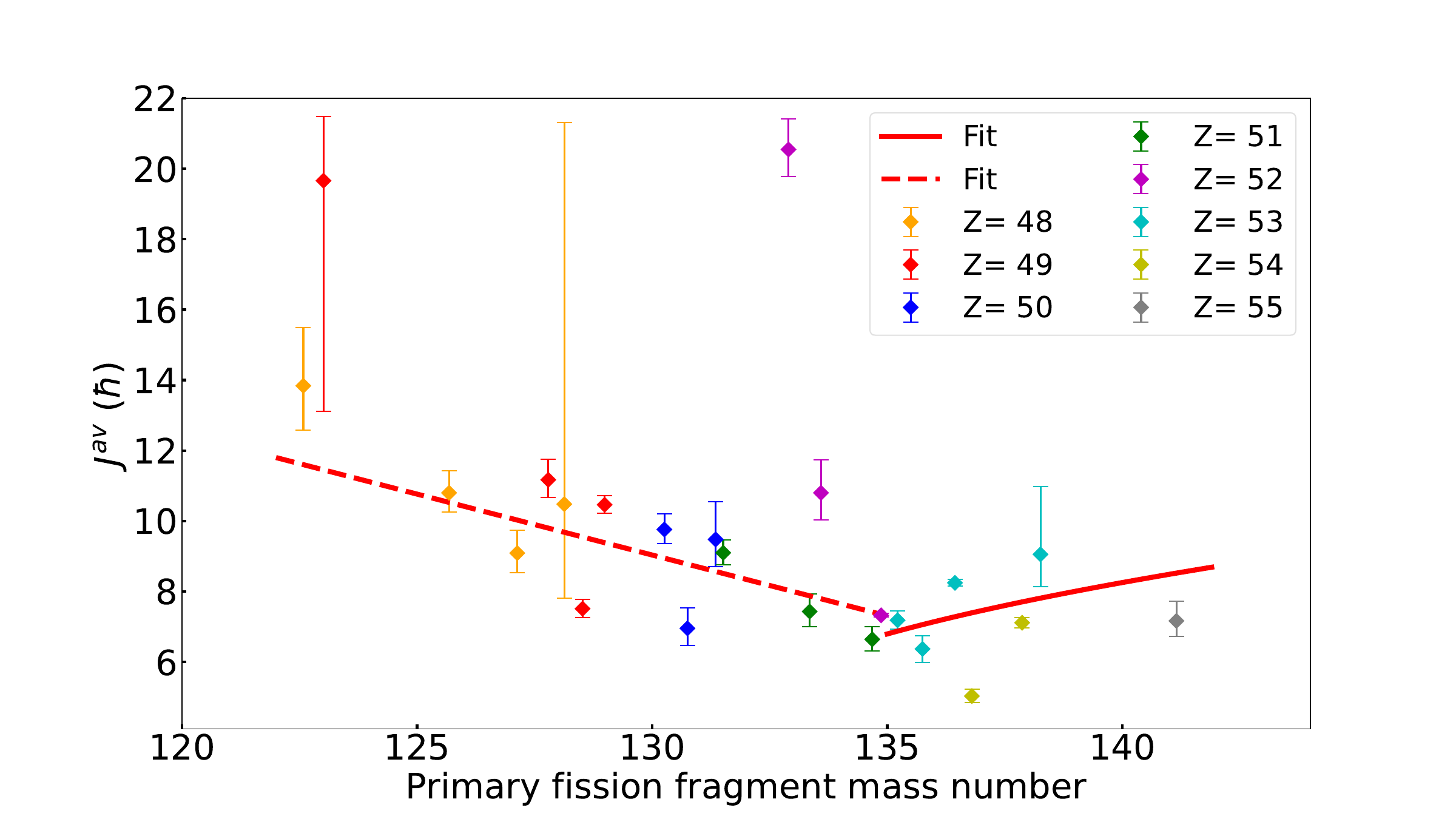}
    \caption{The average angular momentum as a function of the average mass number of the primary FFs de-exciting to the observed FPs. The dashed line represents a linear fit to the angular momentum as a function of the primary FFs in the symmetry region. The red curve is a fit of Eq.~(\ref{J_mass}) to the angular momenta as a function of the primary FFs for A$>$ 134.}
    \label{Jav_FF}
\end{figure}

Considering that the derived angular momentum is a property of the initial fragments, rather than the FPs, it would be interesting to study how this varies with the fragment mass. Using the neutron emission information of the de-excitation from TALYS, shown in Figure~\ref{n_mass}, the average mass number of the FFs de-exciting to a certain FP was calculated. In this way, the fission product mass number in Figure~\ref{J_av} can be shifted to the most probable FF mass number, and the result is presented in Figure~\ref{Jav_FF}. 


Based on the assumption by Wilson~\emph{et al.}, it makes more sense to fit the resulting parametrization (Eq.~(\ref{J_mass})) as a function of the mass of the primary fragments. 
The shift of the minimum of the average angular momenta from 132 to 134 in Figure~\ref{Jav_FF}, implies that the FFs around mass number 134 have the lowest angular momenta. Considering that the neutron multiplicity for FFs around A= 134 in proton-induced fission of $^{238}$U at 20~MeV is about 2~\cite{Kniajeva04}, these FFs will de-excite to FPs around mass number 132.

The decreasing trend of the average angular momentum for A~$<$~134 in Figure~\ref{Jav_FF} is similar to that for A~$<$~132 in Figure~\ref{J_av}. This observation supports our discussion in section~\ref{Compare_J}. \par 


\section{Conclusion and outlook}

A surrogate model of GEF that can be used to generate energy vs angular momentum matrices describing excited FFs has been developed. Using TALYS to calculate the de-excitation of the fragments while varying the parameters of the model, the impact of the state of primary FFs on the IYRs in nuclear fission has been investigated. \par 

31 IYRs in proton-induced fission of $^{238}$U at 25~MeV are compared with the calculated ratios from the model. Based on that comparison, the root-mean-square angular momentum, $J_{rms}$, of 24 FPs have been estimated. When applicable, the estimated $J_{rms}$ are compared with the values derived by V. Rakopoulos~\emph{et al.}, and differences between the two methods are discussed. \par 

It has been shown that the mass dependence of the average angular momentum $J^{av}$ for the nuclides with mass numbers larger than 131 could be reasonably well described by the model suggested by J. Wilson \emph{et al.}~\cite{Wilson21}. This dependence is based on the assumption that the angular momentum is generated from the collective motion of nucleons in the ruptured neck of the fissioning system. \par 

In the symmetric mass region, a decreasing trend of the average angular momentum of the FPs is observed for the first time. This observation could not be explained by the model proposed in Wilson's article and needs a different mechanism to model the angular momentum generation in the symmetry region. \par 

To link the average angular momentum to the primary FFs, the number of emitted neutrons is calculated for each fragment. This way, the average angular momentum as a function of the average mass number of the primary FF is obtained. In this description, the minimum of the angular momentum is close to mass number 134, deviating from the shell closure at 132 by two mass units. \par 

The fact that the surrogate model fails to reproduce the experimental data in some cases shows that the model needs further investigation and development. In particular, the impacts of the (un)known level scheme of the FPs, and possible dependencies of the IYR on parameters other than the angular momentum, are worth future studies. Also, the effect of other parameters in TALYS, such as \textit{Rspincut} possibly related to the momentum of inertia~\cite{talys196}, should be investigated in order to improve the IYR calculation.


\section{Acknowledgement}
This work was supported by the Swedish research council Vetenskapsrådet (Ref. No. 2017-06481), the European Commission within the Seventh Framework Programme through Fission-2013-CHANDA (Project No. 605203), the Swedish Radiation Safety Authority (SSM). The authors would like to acknowledge Harald och Louise Ekmans forskningsstiftelse for the generous grant at Sigtunastiftelsen. \par 

The computations were enabled by resources in project NAISS 2023/22-271 provided by the National Academic Infrastructure for Supercomputing in Sweden (NAISS) at UPPMAX, funded by the Swedish Research Council through grant agreement no. 2022-06725.\par

%
\bibliography{apssamp}

\end{document}